\documentclass[fleqn,usenatbib]{mnras}
\usepackage{newtxtext,newtxmath}
\usepackage[T1]{fontenc}

\usepackage{tikz}
\usepackage{CJK}
\usepackage{multirow}

\usepackage{amsmath, amstext}
\usepackage{graphicx}
\usepackage{enumitem}
\usepackage{natbib}

\newcommand{\amp}{\ensuremath{A}}

\DeclareMathAlphabet\mathbfcal{OMS}{cmsy}{b}{n}

\sloppypar

\definecolor{mygreen}{HTML}{006d2c}
\definecolor{myblue}{HTML}{0868ac}
\definecolor{mybleuclair}{HTML}{f7fbff}
\definecolor{mypurple}{HTML}{6a51a3}
\definecolor{mybrown}{HTML}{67001f}
\definecolor{mypink}{HTML}{980043}
\definecolor{mypinkclair}{HTML}{f7f4f9}
\definecolor{myorange}{HTML}{fe9929}
\definecolor{myorangeclair}{HTML}{fff7ec}
\definecolor{mypurpleclair}{HTML}{fcfbfd}
\definecolor{mybluea}{HTML}{084081}

\title[Unwinding the Snail]{\Large Vertical motion in the Galactic disc: unwinding the Snail}

\author[N. Frankel et al.]{
Neige Frankel,$^{1,2}$\thanks{frankel@cita.utoronto.ca}
Jo Bovy,$^{2}$
Scott Tremaine,$^{1,3}$
David W. Hogg$^{4,5,6}$
\\
$^{1}$Canadian Institute for Theoretical Astrophysics, University of Toronto, 60
St. George Street, Toronto, ON M5S 3H8, Canada\\
$^{2}$David A. Dunlap Department of Astronomy and Astrophysics, University of Toronto, 50 St. George Street, Toronto, ON M5S 3H4, Canada\\
$^{3}$School of Natural Sciences, Institute for Advanced Study, Princeton, NJ 08540, USA\\
$^{4}$Max Planck Institute for Astronomy, K\"onigstuhl 17, D-69117 Heidelberg, Germany\\
$^{5}$Center for Cosmology and Particle Physics, Department of Physics, New York University, 726~Broadway, New~York,~NY 10003, USA\\
$^{6}$Flatiron Institute, 162 Fifth Avenue, New~York,~NY 10010, USA
}

\date{Accepted XXX. Received YYY; in original form ZZZ}

\pubyear{2022}

\begin{document}
\label{firstpage}
\pagerange{\pageref{firstpage}--\pageref{lastpage}}
\maketitle
%
%
%
%
%
%

\begin{abstract}\noindent 
The distribution of stars in the Milky Way disc shows a spiral structure--the Snail--in the space of velocity and position normal to the Galactic mid-plane. The Snail appears as straight lines in the vertical frequency--vertical phase plane when effects from sample selection are removed. Their slope has the dimension of inverse time, with the simplest interpretation being the inverse age of the Snail. Here, we devise and fit a simple model in which the spiral starts as a lopsided perturbation from steady state, that winds up into the present-day morphology. The winding occurs because the vertical frequency decreases with vertical action. We use data from stars in \textsl{Gaia} EDR3 that have measured radial velocities, pruned by simple distance and photometric selection functions. We divide the data into boxels of dynamical invariants (radial action, angular momentum); our model fits the data well in many of the boxels. The model parameters have physical interpretations: one, \amp, is a perturbation amplitude, and one, $t$, is interpretable in the simplest models as the time since the event that caused the Snail. We find trends relating the strength and age to angular momentum: (i)~the amplitude $\amp$ is small at low angular momentum ($<1\,600\mbox{\,kpc\ km\ s}^{-1}$ or guiding-centre radius $< 7.3\,$kpc), and over a factor of three larger, with strong variations, in the outer disc; (ii)~there is no single well-defined perturbation time, with $t$ varying between 0.2 and 0.6\,Gyr. Residuals between the data and the model display systematic trends, implying that the data call for more complex models.
\end{abstract}


%
%
%
%
%
%
\section{Introduction}
Galaxies grow by merging with other galaxies or by forming stars from gas that they accrete \citep[e.g.,][]{rees_ostriker_1977,white_rees_1978,mo_mao_white_1998}. In the course of their formation and evolution, disc galaxies can further develop structures such as bars and spiral arms. These perturbations affect the orbital structure of their host discs by the radial transport of angular momentum and energy \citep{lynden-bell_kalnajs_1972,sellwood_binney_2002,kormendy_2004A}. All these external and internal growth processes can leave distinct dynamical signatures in the dynamics of the host galaxy's stars, dark matter and gas.

In the Milky Way, we expect those perturbations to have only small amplitudes relative to a smooth, axisymmetric state, given its history and semi-isolation. In particular, the Milky Way is thought to have had a relatively calm merger history: its thin disc, with a scale-height of $\sim$ 100--300 pc at the solar radius \citep{bovy_etal_2016,ting_2019}, dubbed the ``low-[$\alpha$/Fe]'' disc, contains stars as old as 8 Gyr, implying that any major merger is older than this. 

The closest luminous satellites of the Milky Way and hence its strongest current external perturbers are the Sagittarius (Sgr) dwarf galaxy \citep{ibata_1994} and the Large Magellanic Cloud, which have masses of $4 \times 10^8 M_\odot$ \citep{vasiliev_belokurov_2020} and $1.4\times10^{11}M_\odot$ \citep{erkal_2019, vasiliev_2021} respectively and distances from the Sun of $\sim 25$ kpc and 50 kpc \citep{ibata_1997, pietrzynski_2013}. 

\begin{figure}
    \centering
    \includegraphics[width=\columnwidth]{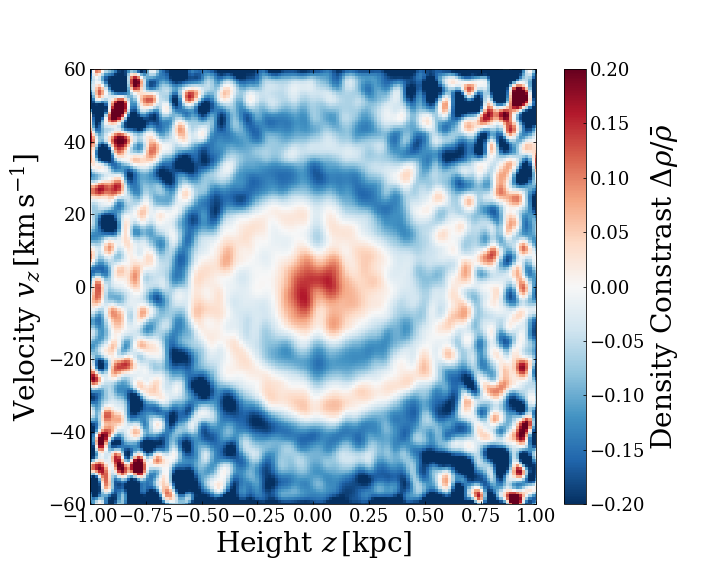}
    \caption{Fractional density contrast in the $z$--$v_z$ plane, revealing the \textsl{Gaia} phase-space Snail. The vertical band near $z=0$ is due to dust extinction. The details of how this figure was constructed are given in \S\ref{section:data_selection}.}
    \label{fig:snail}
\end{figure}

We expect that  disturbances to the vertical equilibrium of the Galactic disc should damp more rapidly than in-plane disturbances, because the orbital frequencies are larger in the direction perpendicular to the mid-plane. Therefore, it was remarkable that \textsl{Gaia} Data Release 2 \citep[DR2,][]{gaia_dr2_2018} revealed a spiral structure in the vertical phase space (position $z$ and velocity $v_z$ normal to the mid-plane) of stars around the Sun \citep{antoja_2018}. This structure, dubbed ``the Snail'' and shown in Figure \ref{fig:snail}, is reminiscent of the vertical waves discovered by \cite{widrow_2012} and analyzed by \cite{widrow_2014}, although these papers focused on the vertical motion of the disc as a whole rather than its internal structure. The vertical waves, well visible in \textsl{Gaia} DR2 \citep{bennett_bovy_2019}, may also be related to the galactic warp \citep{poggio_2021} and large-scale corrugations \citep{xu_2015}.
The Snail was originally seen most prominently in the mean azimuthal velocity as a function of $z$ and $v_z$ but is also clear in plots of the fractional density contrast relative to a smooth distribution (Fig. \ref{fig:snail}). Such a spiral in phase space is almost certainly the signature of an on-going phase mixing process.

The Snail could have emerged via various mechanisms. The simplest of these involves a single close encounter with a massive perturber external to the disc, presumably a satellite galaxy or a dark-matter subhalo. In these models, the degree of winding of the snail provides a direct measure of the encounter time. The derived encounter time is consistent with the orbit of the Sgr dwarf galaxy, although with large uncertainties  \citep[e.g.,][]{antoja_2018}. Simulations exploring this scenario \citep{laporte_2018_sgr,laporte_2019_sgrfootprint,bland-hawthorn_2019, bennett_bovy_2021,hunt_2021, bennett_2022, gandhi_2022_snail}, ranging from test-particle simulations to idealized N-body simulations to cosmological simulations, have shown that the perturber mass necessary to reproduce the amplitude of the observed Snail is larger than the current Sgr mass by 
factors of a few: the present-day dynamical mass of the Sgr dwarf is $\sim 4\times 10^8 M_\odot$ \citep{vasiliev_belokurov_2020}, whereas these simulations, including the stripping of the dwarf as it falls in, require a present-day mass of $\sim10^{9}$--$3\times 10^{10}M_\odot$ or very rapid mass loss in the recent past. It is unlikely that the Snail arises from a single encounter with a different known satellite, because they all produce weaker responses by an order of magnitude or more \citep[][Fig. 7]{banik_2022}. It is also unlikely that the Snail is excited by an invisible dark-matter subhalo, since subhalos massive enough to excite the Snail should also be massive enough to form and retain a substantial population of stars.  

Another hypothesis is that the Snail is excited by the buckling of the Galactic bar \citep{khoperskov_2019}, an event that can produce a Snail with the observed amplitude. However, there are concerns with this scenario: (i) It is likely that any one-time buckling event occurred not long after the disc was formed, and a Snail cannot survive for much longer than a Gyr \citep{tre22}. (ii) It is surprising that the Snail in this simulation seems clearer when color-coded by radial velocity rather than by density contrast, which is not the case in the \textsl{Gaia} data \citep{ls20}. (iii) Other realistic simulations of a Milky Way-like barred galaxy do not exhibit a Snail-like feature in the present solar neighbourhood \citep{tepper21}.

High-resolution cosmological simulations can produce phase-space spirals in galactic discs even when there are no massive nearby satellites to excite them  \citep{garcia_conde_2022}, so other processes such as resonances or non-axisymmetric structures in the disc \citep{khoperskov_2019}, spatially or temporally inhomogeneous star formation, halo wakes \citep{grand_2022}, or the cumulative effects of many low-mass subhalos \citep{tre22} might also be responsible for the Snail.  

Here, we set out to construct a simple parametric model for the Snail, and to fit the model parameters to data from \textsl{Gaia} EDR3. We split the stars in the sample depending on their actions in the disc plane ($J_\varphi, J_R$), and extract the snail amplitude and degree of winding as a function of the actions. Although this simple model does not capture all the physics participating in the Snail's creation and evolution (and is not meant for this), it can be straightforwardly applied to any simulation output and therefore can be used to link quantitatively the snails in simulations to those in the Milky Way. 

In \S\ref{section:freq-angle}, we unwind the Snail by transforming the stellar distribution from position vs.\ velocity phase space to vertical frequency vs.\ angle space, the $\Omega_z$--$\theta_z$ plane. This transformation requires us to correct for selection effects, which severely affect the angle distribution. The frequency-angle plane has the advantage that the arms of the Snail should appear as straight lines if they are created by a single instantaneous event in the distant past, whether a satellite encounter, bar buckling, or something else (\citealt{tre22} argue that even some steady-state processes can also produce approximately straight features in this plane). In \S\ref{section:model}, we construct and fit a simple model for the distribution of stars in the vertical frequency--angle plane; the fits are carried out both in angular-momentum bins and in boxels of angular momentum and radial action. 
We present the results in \S\ref{section:results}. In \S\ref{section:discussion} we discuss the results and describe the limitations and extensions of our models.

%
%
%
%
%
%

\section{The Snail in the Frequency-Angle Plane} \label{section:freq-angle}

In this Section, we apply the data transformations required to unwind the Snail. We correct for selection effects and show that the Snail appears as a set of nearly straight stripes in the $\Omega_z$--$\theta_z$ plane, which motivates our subsequent modelling.

\subsection{Data selection and products}\label{section:data_selection}
\begin{figure}
    \centering
    \includegraphics[width=\columnwidth]{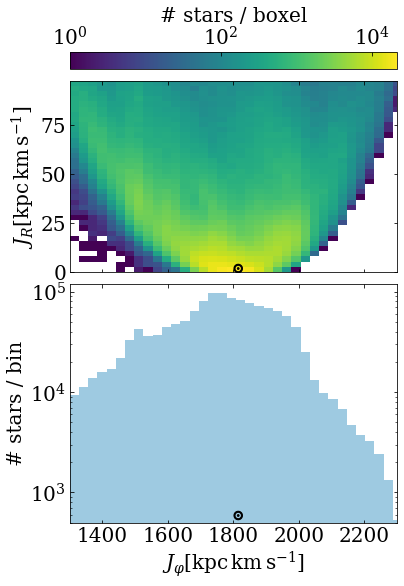}
    \caption{Distributions of orbits of the stars in our sample. The $\odot$ symbol marks the angular momentum of a circular orbit at the solar radius, $J_{\varphi\odot}=1\,811\,\mathrm{kpc\,km\,s^{-1}}$.}
    \label{fig:action_distributions}
\end{figure}

\begin{figure}
    \centering
    \includegraphics[width=\columnwidth]{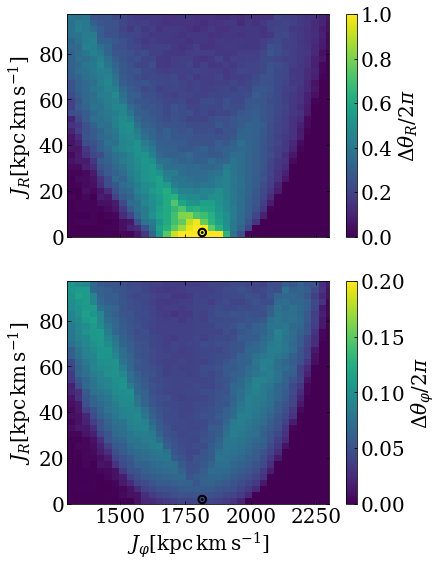}
    \caption{Fraction of the range of orbital angle covered by the data set. The top panel shows the radial angle $\theta_R$ and the bottom panel shows the azimuthal angle $\theta_\varphi$. The plotted quantities $\Delta\theta_{R,\varphi}$ are evaluated by binning the stars in angles from 0 to $2\pi$ and counting the fraction of non-empty bins. The $\odot$ symbol marks the angular momentum of a circular orbit at the solar radius.}
    \label{fig:angle_coverage}
\end{figure}
%
%

We base our analysis on the \textsl{Gaia} EDR3 sub-sample with radial velocities \citep[the \textsl{Gaia} RVS sample,][]{gaia_prusti_2016, Gaia_EDR3_Brown_2021} using the ADQL query in Appendix \ref{appendix:data_query}. We have performed only simple colour and magnitude cuts  ($3.4 < G < 12.5$ and $0.35 < G-G_\mathrm{RP} < 1.25$) in order to produce a sub-sample with a well-defined selection function in phase space, and selected the stars inside a cylinder centered on the Sun with a radius of 0.5 kpc. The quartiles of the distance distribution are $0.23$, $0.33$ and $0.44$ kpc. 

The colour cuts  follow the recommendations of \cite{rybizki_2021}, who point out that molecular bands in cold stars ($T_\mathrm{eff} \leq 3550$ K) and strong Paschen lines in hot stars ($T_\mathrm{eff} \geq 6900$ K) can prevent an accurate measurement of the line-of-sight velocity. These temperature cuts approximately correspond to colours of $G-G_\mathrm{RP}= 0.35$ and 1.25. Between these colours, the completeness of the RVS sample should be the same as that of the underlying astrometric sample \citep{rybizki_2021}. We make the two magnitude cuts at the bright and faint end to have control over the magnitude range of stars in the sample rather than having them be determined by, e.g., instrumental effects.

These cuts produce a sample of $1,402,553$
stars with complete six-dimensional phase-space coordinates. We experimented with additional cuts based on the fractional parallax uncertainty but these did not change our results substantially. 

To convert to Galactocentric coordinates, we assume that the distance to the Galactic center is $R_\odot = 8.23\,\mathrm{kpc}$ \citep{leung_2022}, that the Sun's height with respect to the Galactic mid-plane is $z_\odot = 20.8$ pc \citep{bennett_bovy_2019}, and that the solar motion with respect to the Local Standard of Rest is  $v_\odot = (11.1, 12.24, 7.25)\mbox{\,km s}^{-1}$ \citep{schoenrich_2010}. We use the default circular velocity at the solar radius implemented in the \textsc{galpy} software package\footnote{\url{https://github.com/jobovy/galpy}~.} \citep{bovy_2015_galpy}, $220\,\mathrm{km\,s}^{-1}$. We compute the height and velocity of each star normal to the Galactic mid-plane, $z$ and $v_z$, as well as the Galactocentric distance $R$ and the $z$-component of the angular momentum, $J_\varphi=R\,v_\varphi$ where $v_\varphi$ is the azimuthal velocity in inertial Galactocentric coordinates. Using the Milky Way potential \textsc{MWPotential2014} from \textsc{galpy}\footnote{We have tested the \citet{mcmillan_2017} potential as well and found similar results; see \S\ref{section:discussion}.}, we calculate actions $(J_R,J_\varphi,J_z)$, angles $(\theta_R,\theta_\varphi,\theta_z)$, and frequencies $(\Omega_R,\Omega_\varphi,\Omega_z)$ using the Staeckel Fudge \citep{binney_2012} as implemented in \textsc{galpy}. In particular the vertical action and frequency are defined as
\begin{equation}
    J_z=\frac{1}{2\pi}\oint v_z\,dz, \quad \Omega_z=2\pi\left(\oint\frac{dz}{v_z}\right)^{-1},
\end{equation}
where the integral is over one complete orbit in $z$. 

The resulting distribution of orbits is shown in Fig.\ \ref{fig:action_distributions}, as a function of angular momentum and radial action ($J_\varphi, J_R$) in the top panel, and as a function of angular momentum ($J_\varphi$) in the bottom panel.
In Fig.\ \ref{fig:angle_coverage}, we show the fraction of radial and azimuthal angles that are covered by the data on the $J_\varphi$--$J_R$ plane (which are mostly set by our spatial selection, the magnitude of the brightest stars in the sample, and our colour-magnitude cuts).  In the regions where this fraction is less than 1, only some orbital phases are represented by the data set.  In the top panel, we show the radial angle $\theta_R$. In the region of low radial action $J_R$ near the angular momentum of a circular orbit at the solar radius, $J_{\varphi\odot} = 1\,811\mathrm{\,kpc\,km\,s^{-1}}$, all radial phases are sampled, i.e., we see stars from apocenter to pericenter.   At low angular momentum, we only see stars close to their apocenters whereas at high angular momentum, we mostly see stars near their pericenters. In the bottom panel, we show the azimuthal angle $\theta_\varphi$; at best 15\% of the azimuthal angle is covered. These trends are all expected for a sample localized around the Sun. 

Using this data set, we show the density contrast $\Delta \rho/\bar{\rho}=\rho/\bar{\rho}-1$ in the $z$--$v_z$ plane in Fig.\ \ref{fig:snail}. The phase-space density ${\rho}(z,v_z)$ is obtained through Gaussian kernel density estimation using the \textsc{scipy} sofware package \citep{scipy}. The covariance matrix of the Gaussian kernel is that of the data set multiplied by Scott's factor \citep{scott_KDEBandwidth}. The mean phase-space density $\bar{\rho}$ is computed by smoothing $\rho$ with a Gaussian filter with a scale of 40 pc in $z$ and $5\,\mathrm{km\,s^{-1}}$ in $v_z$.

The phase-space snail in Fig.\ \ref{fig:snail} can be traced over at least two revolutions. The direction of the spiral--counterclockwise as one travels outward from the origin at $z=v_z=0$--is consistent with phase wrapping of an initial perturbation without spirality, since the frequency of the vertical oscillations ($\Omega_z$, defined below)  is a decreasing function of the vertical amplitude or action in any realistic Galactic potential.

%
%
\subsection{Selection correction \label{section:selection_function_correction}}

To unravel the Snail in the frequency-angle plane, we first need to correct for selection effects arising from our cuts in apparent magnitude (see  \S\ref{section:data_selection}). 
\begin{figure*}
    \centering
    \includegraphics[width=\textwidth]{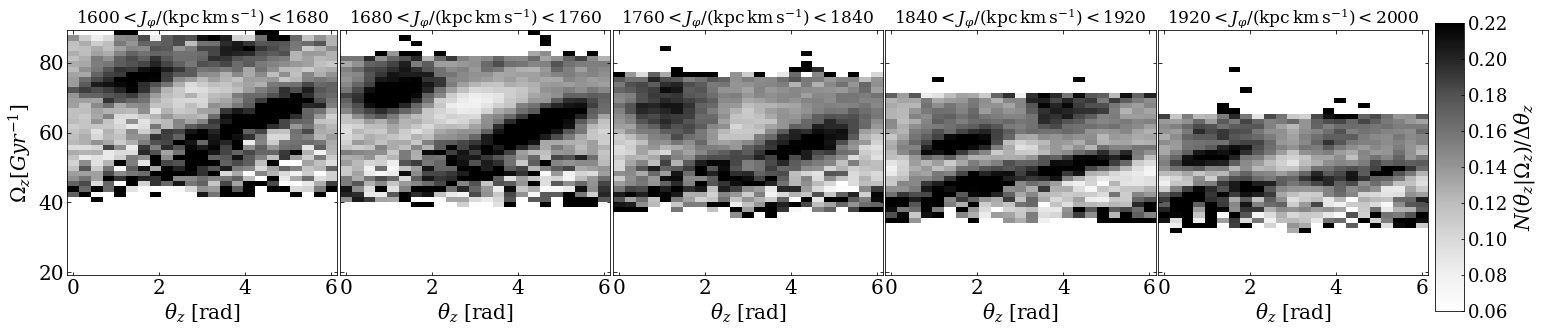}
    \caption{Signatures of unmixed stellar populations in the vertical frequency--vertical phase plane with coordinates $(\Omega_z, \theta_z)$ (the zebra diagram). This is a row-normalized and selection-reweighted histogram. It represents the number count of points in a $\Omega_z$--$\theta_z$ boxel divided by the total number of points at given frequency $\Omega_z$, and thereby represents the distribution $p(\theta_z|\Omega_z, J_\varphi)$.
    The five panels show adjacent intervals of the $z$-component of angular momentum, $J_\varphi$. The angular momentum of a circular orbit at the solar radius is $J_{\varphi\odot} = 1\,811\,\mathrm{kpc\,km\,s^{-1}}$.}
    \label{fig:zebra}
\end{figure*}

\begin{figure*}

    \centering
    \includegraphics[width=\textwidth]{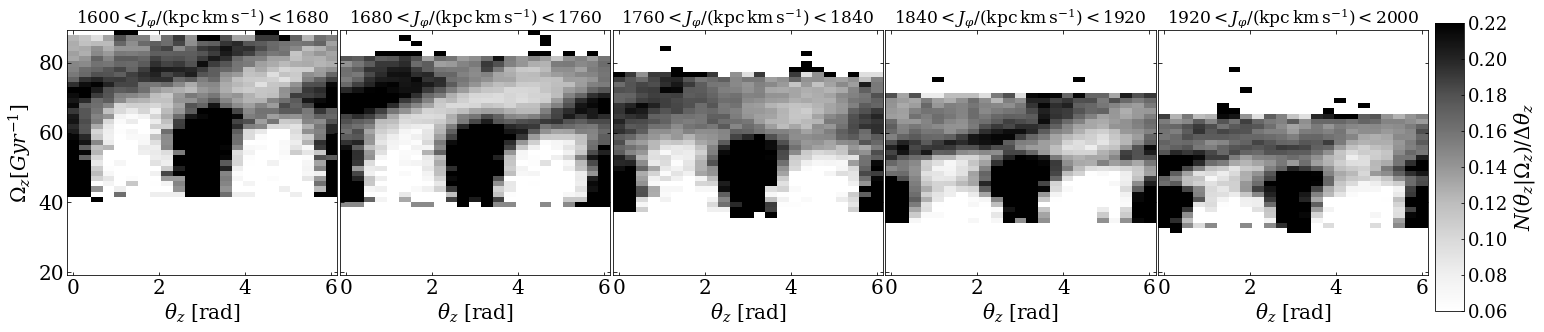}
    \caption{Same as Fig. \ref{fig:zebra}, without applying the correction for selection effects, i.e., these panels represent the distribution $p(\theta_z|\Omega_z, J_\varphi)$ in the data set rather than the underlying population. The vertical stripes, representing over-densities at $\theta_z = 0$ and $\pi$, arise at low frequency (i.e., large vertical action, amplitude or energy) because a fraction of stars with large vertical amplitudes are too faint (too distant) to survive the selection cut on apparent magnitude.}
    \label{fig:zebra_nocorrec}
\end{figure*}

\subsubsection{Modelling the selection}
We define two distribution functions of the actions and angles. The first, $p_\mathrm{MW}(J_R,J_\varphi,J_z,\theta_R,\theta_\varphi,\theta_z)=p_\mathrm{MW}(\vec{J},\vec{\theta})$, is defined such that $p_\mathrm{MW}d\vec{J}d\vec{\theta}$ is the probability that a star in the Milky Way lies in the phase-space volume element $d\vec{J}d\vec{\theta}$ at the present time. The second is the observed distribution in the vertical action and angle, $p_\mathrm{O}(J_z, \theta_z |J_R,J_\varphi,\theta_R,\theta_\varphi)$,
defined such that $p_\mathrm{O} dJ_z d\theta_z$ is the probability that a star in our sample with the given values of $J_R$, $J_\varphi$, $\theta_R$, and $\theta_\varphi$ lies in the small volume element $dJ_z d\theta_z$.

To model the effects of selection, we assume \textit{for the purpose at hand} that interstellar extinction is negligible in the cylinder of stars containing our sample within a projected distance onto the Galactic mid-plane of 0.5 kpc from the Sun. This is true for most portions of the sky, although there are a few small angular regions with extinction values as high as $A_G=2$ magnitudes (dust is responsible for the thin vertical band at $z=0$ seen in Fig.\ \ref{fig:snail}). Fortunately, most dust-related effects are strongest at the mid-plane, and therefore appear as even harmonics in the angle $\theta_z$, as can be seen in Fig.\ \ref{fig:zebra} where there are vertical under-density bands when stars cross the mid-plane, i.e., at $\theta_z=0$ going upwards and $\theta_z = \pi$ going downwards. However, since in the solar neighbourhood the Snail is strongest in the $m=1$ harmonic, the effects of dust probably do not compromise our conclusions about the Snail.

With these assumptions, the apparent magnitude of a star in the \textsl{Gaia} $G$-band depends only on its absolute magnitude $M_G$ and its distance $D$. We assume that the probability $S(M_G,D)$ that a star with colour in the range $0.35 < G-G_\mathrm{RP} < 1.25$ will appear in our sample is

\begin{equation}
S(M_G, D) = C_\mathrm{norm} \left[ S_1(M_G, D) + S_2 \right]
\end{equation}
where 
\begin{equation}
    S_1(M_G,D)=\left\{\begin{array}{ll} 1, & 3.4 < M_G+5\log_{10}D/10\mathrm{\, pc}<12.5 \\
    0, & \mathrm{otherwise}
    \end{array}\right.  
    \label{eq:sdef}
\end{equation}
and 
\begin{equation}
    S_2 = \mathrm{offset} \ll 1\,.
    \label{eq:sdef1}
\end{equation}
We introduced the offset $S_2$ because otherwise the corrections for selection effects described below will sometimes lead to division by small numbers, and thus to unrealistically large weights for a small number of stars. This can occur in regions of the parameter space that only have a few data points, or if there are features of the data that are not included in the simple form of Eq.\ (\ref{eq:sdef}). For example, we obtain the distance $D$ to a star by inverting the parallax $\varpi=1/D$ and we do not account for parallax uncertainties; thus there could be some stars with a measured distance $D$ that places them in a region of the parameter space where the modelled probability for them to be selected is unrealistically small. Here, we set $S_2 = 10^{-5}$, implying that only a fraction $f_w < 6\times10^{-6}$ of the sample (corresponding to distances $D\gtrsim 6\,$kpc) would have weights $w(D)$ as defined in Eq.\ (\ref{eq:wdef}) in which the contribution from $S_1$ is less than the contribution from $S_2$. We have also explored two larger offsets, $S_2 = 0.005$ and $S_2 = 0.02$ (i.e., $f_w \simeq 2\times 10^{-4}$ corresponding to a distance $D\gtrsim3\,$kpc, and $f_w \simeq 8\times 10^{-3}$ corresponding to a distance $D\gtrsim1\,$kpc), and found quantitatively similar results. Physically, we know that the Snail is most easily detected in stars within about 1 kpc, and the extra term $S_2$ penalizes stars at much larger distances such that they do not contribute too much to our sample due to their low probability of selection. The constant $C_\mathrm{norm}$ can be chosen to be $(1+S_2)^{-1}$, such that the maximum value of the probability $S(M_G,D)$ is unity, but in fact all the calculations below are independent of the value of $C_\mathrm{norm}$.

Using the selection probability $S(M_G,D)$, we can relate  $p_\mathrm{O}(J_z,\theta_z|J_R,J_\varphi,\theta_R,\theta_\varphi)$ to $p_\mathrm{MW}(\vec{J},\vec{\theta})S(M_G,D)$ as
\begin{align}
    &p_\mathrm{O}(J_z,\theta_z|J_R,J_\varphi,\theta_R,\theta_\varphi)\nonumber \\
    &= c\int p_\mathrm{MW}(\vec{J},\vec{\theta})S(M_G,D)p(M_G)dM_G \\
    &= c \times p_\mathrm{MW}(\vec{J},\vec{\theta})w(D),
\label{eq:popmw}
\end{align}
where
\begin{equation}
    w(D)\equiv \int p(M_G)S(M_G,D)dM_G.
    \label{eq:wdef}
\end{equation}
Here the distance $D(\vec{J},\vec{\theta})$ is a function of the actions and angles, determined by the gravitational potential; $p(M_G)$ is the luminosity function of stars satisfying the colour cut, normalized such that $\int p(M_G)dM_G=1$; the  luminosity function is assumed to be independent of position; and $c(J_R,J_\varphi,\theta_R,\theta_\varphi)$ is a normalizing constant determined by the condition that $\int p_\mathrm{O} dJ_z d\theta_z=1$.

\subsubsection{Nuisance variables: modelling the luminosity function}
We now build an empirical model of the luminosity function $p(M_G)$ from nearby stars. 
We select a ``local'' sub-sample of stars from our sample having  distance $D<100\,$pc from the Sun and assume that in such a small volume, the spatial distribution is roughly uniform: the strongest density variations are those with height $z$, but the population-averaged local scale-height is $h_z \sim 300\,$pc $> 100\,$pc \citep{bovy_etal_2012}; numerical experiments show that  accounting for those vertical variations would lead to the same weight values to within 15\%. The luminosity function describes the relative number of stars of different luminosities (here, magnitudes). The fraction of stars in the solar neighbourhood with magnitude $M_G\pm \Delta M_G/2$ is $p(M_G)\Delta M_G$. It can be obtained from the number of stars in the sample in the same magnitude range $\Delta N$, divided by their observable volume $V(M_G)$ given their magnitude: $p(M_G)\Delta M_G \propto \Delta N/V(M_G)$. Here $V(M_G)=(4\pi/3) D_\mathrm{max}(M_G)^3$, with $D_\mathrm{max}(M_G)=\min[100\,\mathrm{pc}, D_{12.5}(M_G)]$ and $D_{12.5}=0.01\times 10^{(12.5 - M_G)/5}$ kpc, corresponding to the distance at which a star with absolute magnitude $M_G$ becomes fainter than the flux limit of our sample, $G=12.5$.
 In practice, we do not need to bin the luminosity function: instead we use bootstrap resampling with a probability $\propto 1/V(M_G)$ to create a set of stars that is sampled from the luminosity function. The integral over $M_G$ in Eq.\ (\ref{eq:wdef}), leading to $w(D)$, can be evaluated directly from this set of stars.

\subsection{Unwinding the Snail into the `zebra diagram'}
\label{section:unwindingsnail-zebra}
At this point, we could forward-model the data by constructing a parametrized form for $p_\mathrm{MW}(\vec{J},\vec{\theta})$, converting it to the observed distribution $p_\mathrm{O}(J_z,\theta_z|J_R,J_\varphi,\theta_R,\theta_\varphi)$ using Eq.\ (\ref{eq:popmw}), and adjusting the parameters to find the best fit to our sample. We prefer instead to  ``correct'' the observed distribution $p_\mathrm{O}$ to determine $p_\mathrm{MW}$, by dividing it by $w(D)$. In practice, we use bootstrap resampling (i.e., resampling with replacement) of stars from the data set with a probability $P = 1/[w(D)\sum_{i}1/w(D_i)]$, to increase the weight of data that had a low probability to be selected in our data set; we call this `selection reweighting'. We set the reweighted sample size to be the same as the original sample size.

We show in Fig.\ \ref{fig:zebra} a row-normalized histogram of the stars in the frequency-angle or $\Omega_z$--$\theta_z$ plane. The histogram is selection reweighted (see Fig. \ref{fig:zebra_nocorrec} for the same histogram without selection reweighting). The angled stripes in this ``zebra diagram" are a manifestation of the \textsl{Gaia} Snail \citep[a similar plot is shown by][]{li_widrow_2021}. The stripes have a simple physical interpretation.  Suppose that an instantaneous event in the past caused a perturbation in the distribution function of the form $\Delta p(J_z,\theta_z)=A\cos [m(\theta_z-\theta_{z0})]$. In the unperturbed potential, the action $J_z$ is constant and the angles increase at a rate $\dot\theta_z=\Omega_z$. Therefore at the present time the perturbation has the form $\Delta p(J_z,\theta_z)=A\cos [m(\theta_z-\Omega_zt-\theta_{z0})]$ where $t>0$ is the time elapsed since the event. The extrema of this function are straight lines in the $(\theta_z,\Omega_z)$ plane with slope $1/t$. The features in Fig.\ \ref{fig:zebra} are not expected to be exactly straight and parallel, for several reasons, including the following: (i) the amplitude $A$ and the phase $\theta_{z0}$ may depend on $J_z$; (ii) there may be differences between the assumed gravitational potential and the real potential of the Milky Way; (iii) more than one event may contribute to the Snail. 

An example of systematic deviations from this simple model is that the panels in Fig.\ \ref{fig:zebra} representing stars with smaller angular momentum $J_\varphi$ appear to have steeper slopes (smaller age $t$) than the ones representing high angular-momentum stars, a feature that we explore quantitatively below.

%
%
\section{Modelling the Zebra Diagram} \label{section:model}

In this section, we explore the hypothesis that the \textsl{Gaia} Snail arises from the phase mixing of a single disturbance to the nearby Galactic disc. The most general disturbance of this kind to the distribution function can be written as a sum of terms of the form $A_m\cos[m(\theta_z-\Omega_zt-\theta_{z0,m})]$ where $m\ge 0$ is an integer and the amplitudes $A_m$ and phases $\theta_{z0,m}$ can depend on the actions $J_R,J_\varphi,J_z$ and the angles $\theta_R,\theta_\varphi$. Since the origin(s) of the perturbation is (are) unknown and the amplitudes and phases can have a complex dependence on the stellar orbits, we do not parametrize the model as a function of all of these variables. Instead, we (i) ignore the dependence of the amplitudes and phases on  $J_R$ and $\theta_R$, since the radial excursions of most disc stars are small, and the vertical frequency $\Omega_z$ is not a strong function of $J_R$ (we relax the latter assumption in \S\ref{section:snail_JphiJR}); (ii) consider only $m=1$, since this is the dominant wave number for perturbations from dark-matter substructure or dwarf galaxies \citep{banik_2022}; (iii) ignore the dependence of the amplitudes and phases on $\theta_\varphi$ since we can only sample a limited range of azimuths (see bottom panel of Fig.\ \ref{fig:angle_coverage}). We therefore fit the model to the data from several groups of stars binned by the $z$-component of angular momentum $J_\varphi$\footnote{For reference, at the solar radius $R_\odot=8.23\,$kpc, a circular orbit has $J_\varphi=J_{\varphi\odot}=1\,811\,\mathrm{kpc\,km\,s^{-1}}$.}. Stars that are in our sample and have similar angular momenta should be at roughly the same azimuth at the time when they were perturbed if the perturbation is not too far in the past, since $\Omega_\varphi$ depends mostly on $J_\varphi$ rather than the other actions. 

In each angular-momentum bin, we fit three model parameters $\{\amp, t, \theta_{z0}\}$, using the formula
\begin{equation}\label{eq:model_ml}
\begin{split}
    p_\mathrm{MW}&(J_z, \theta_z|J_\varphi, J_R) = \frac{1}{2\pi}p_0(J_{z}) \\
    &\times \left\{1 + \amp \cos[\theta_z - \Omega_z(J_R, J_\varphi, J_z)t - \theta_{z0}]\right\}\,,
\end{split}
\end{equation}
where $p_0(J_z)$ is the undisturbed distribution of vertical actions, $\amp$ is the strength of the on-going phase-mixing signal, and $\theta_{z0}$ is the phase, that is, the angular symmetry axis of the perturbed distribution at the time of the perturbation event. 
The degree of winding of the Snail is described by the parameter $t$, which has dimensions of time and represents the time elapsed since the perturbation event.

\begin{figure}
    \centering
    \includegraphics[width=\columnwidth]{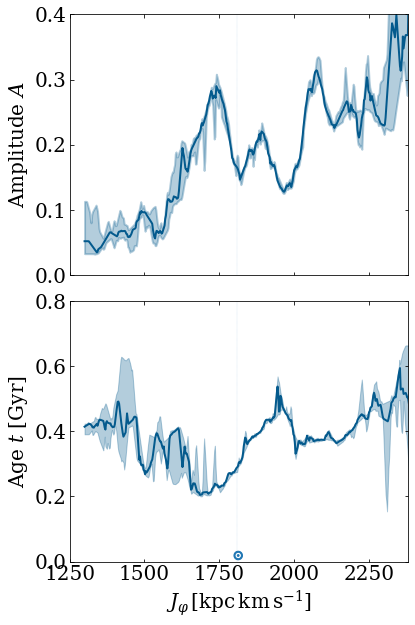}
    \caption{Maximum-likelihood estimates of two of the parameters in Eq.\ (\ref{eq:logl}): the strength of the perturbation, $\amp$ (top panel) and the time of the perturbation, $t$ (bottom). Each solid line corresponds to the median of many different fitting experiments, as specified in the text, in which we varied the initial guess for the optimizer, the binning in $J_\varphi$ and the random subset bootstrapped from the data. The shaded areas enclose the 25$^\mathrm{th}$ and the 75$^\mathrm{th}$ percentiles, as determined from these experiments. The $\odot$ symbol marks the angular momentum of a circular orbit at the solar radius.}
    \label{fig:best_fit_trends}
\end{figure}

\begin{figure*}
    \centering
    \includegraphics[width=\textwidth]{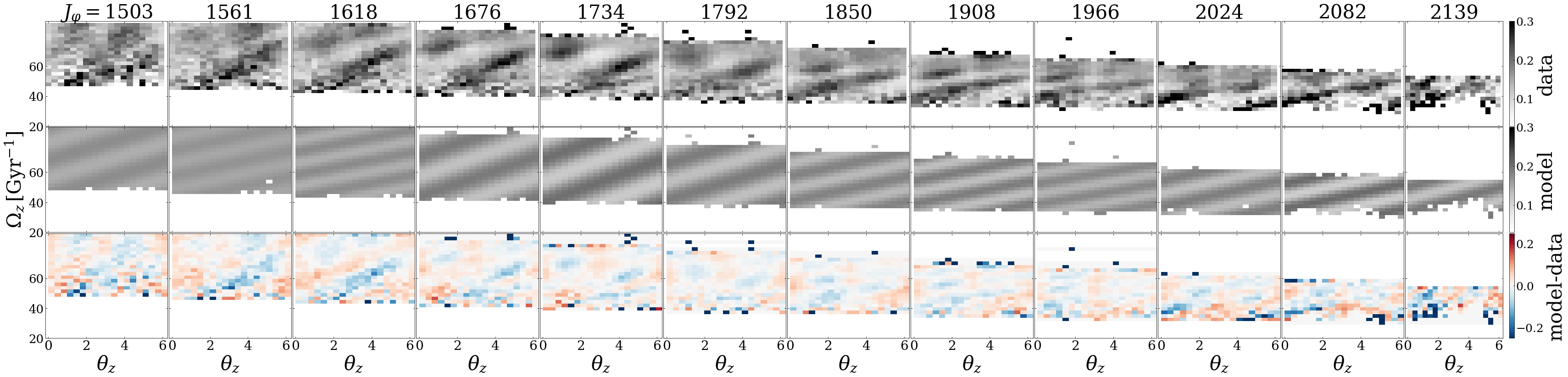}
    \caption{A row-normalized histogram of the number of stars as a function of angle $\theta_z$ at given vertical frequency $\Omega_z$ for 12 bins in angular momentum $J_\varphi$ ($\mathrm{kpc\,km\,s^{-1}}$), i.e., $p(\theta_z|\Omega_z)$. The lines of panels show the data (top), the best fit from the model of Eq.\ (\ref{eq:logl}) (middle), and the residuals between the data and this model (bottom).}
    \label{fig:best_fit_trends_res}
\end{figure*}

The joint distribution of the $N_\star$ selection-reweighted stellar orbits is the product leading to the likelihood function of the model parameters $(\amp, t, \theta_{z0})$ given the selection-reweighted data $\mathcal{D} = \{(\vec{J}_i, \vec{\theta}_i)\}_{i=1,\ldots,N_\star}$:
\begin{equation}
    \mathcal{L}(\amp, t,\theta_{z0};\mathcal{D}) = \prod_{i=1}^{N_\star} p_\mathrm{MW}(J_{zi}, \theta_{zi}|J_{\varphi i}, J_{Ri}) 
\end{equation}
The log-likelihood becomes
\begin{equation}
    \ln(\mathcal{L}) = \sum_{i=1}^{N_\star} \ln \left[1 + \amp\cos\left(\theta_z - \Omega_zt - \theta_{z0}\right)\right] 
    \label{eq:logl}
\end{equation}
plus unimportant terms that are independent of the fitting parameters. 
In practice, since the cosine function is non-linear and periodic, we do not fit directly for $\amp$ and $\theta_{z0}$ but rather write $\amp \cos(\theta_z - \Omega_zt - \theta_{z0}) = a\cos(\theta_z - \Omega_zt) + b \sin(\theta_z - \Omega_zt)$ and fit for $a$ and $b$, with $\amp = \sqrt{a^2 + b^2}$ and $\theta_{z0}=\mbox{arctan2}(b,a)$. We optimize the log-likelihood numerically with the \textsc{scipy} implementation of the Nelder--Mead algorithm \citep{scipy, NeldMead65} and we verify visually that this best-fit model is a plausible fit to the data.

\vspace{-0.5mm}
%
%
\section{Results from the model fit}\label{section:results}

Here we present the best-fit models of the Snail. In \S\ref{section:snail_Jphi}, we extract Snail parameters in bins of $J_\varphi$ and in \S \ref{section:snail_JphiJR} we divide the data into boxels of the dynamical invariants $(J_\varphi, J_R)$.

\subsection{Snail as a function of $J_\varphi$}\label{section:snail_Jphi}

In Fig.\ \ref{fig:best_fit_trends}, we show the best-fit parameters $\amp$ (amplitude) and $t$ (age) as a function of angular momentum. We show the median (solid line) and quartiles around it (shaded region) of 17 fitting experiments. Each of these 17 experiments differs in the initial guess for the time $t$ of the event (0.3, 0.35, 0.4, 0.45 Gyr), in the binning in $J_\varphi$ (20, 40, 80, 100 bins), and in the random samples bootstrapped from the parent data set to correct for selection effects. The experiment with 100 bins in $J_\varphi$ used bin sizes that were adapted such that each bin contains roughly the same number of data points, rather than the fixed bin sizes that we use otherwise. 
The lines plotted in this figure are summary statistics of all these experiments.

We make several remarks about this figure. (i) The bins with $J_\varphi\gtrsim\, 2\,200\,\mathrm{kpc\ km\, s}^{-1}$ contain very few stars, as shown in Fig.\ \ref{fig:action_distributions}, and therefore have a larger scatter in the best-fit amplitude and timescale. (ii) The amplitude $\amp$ is significantly smaller for stars with $J_\varphi\lesssim 1\,600\,\mathrm{kpc\ km\, s}^{-1}$, corresponding to a mean Galactocentric radius $\lesssim 7.2\mbox{\,kpc}$: $\amp\simeq 0.05$ compared to $\gtrsim 0.2$ at higher angular momentum. This sharp change might arise because stars inside $7.2\mbox{\,kpc}$ are sufficiently far from the closest approach of the hypothetical impactor that excited the snail that the vertical components of their orbits were adiabatically invariant during the encounter. (iii) The dynamical age of the event is between 0.2 and 0.6 Gyr in the past; the rather large variations with angular momentum suggest that models in which the Snail arises from a single event of short duration may be oversimplified.

In Fig.\ \ref{fig:best_fit_trends_res}, we display the zebra diagrams of the data, one of the best-fit models, and their residuals. The bins with the smallest angular momentum $J_\varphi$, which have the lowest amplitudes $\amp$, also show the strongest residuals. These residuals have the signature of an $m=2$ mode (particularly in the left-most panel), reminiscent of the breathing or $m=2$ mode described by \citet{hunt_2022}, suggesting that at low angular momentum ($J_\varphi \simeq1\,500\,\mathrm{kpc\ km\,s}^{-1}$) the Snail would be better described by a two-armed spiral. 

In all of the angular-momentum bins, the residuals show structure rather than Poisson noise, implying that the model in Eq.\ (\ref{eq:model_ml}) is overly simple: the amplitude or phase could be a function of vertical action, or the Snail could be excited by multiple impacts at different times $t_i$.

The parameters used to produce the diagrams of Fig.\ \ref{fig:best_fit_trends_res} are documented in Table \ref{table:best_fit}. The initial phases $\theta_{z0}$ exhibit so much scatter that they are almost meaningless; this is mostly because of the strong covariance between $\theta_{z0}$ and $t$.

\begin{table}
\centering
\caption{Best-fit Snail parameters as a function of $J_\varphi$, for the models plotted in Figure \ref{fig:best_fit_trends_res}. The angular momentum $J_\varphi$ is in units of $\mathrm{kpc\,km\,s^{-1}}$, $t$ is in $\mathrm{Gyr}$ and $\theta_{z0}$ is in radians.}\label{table:best_fit}
\begin{tabular}{rrrr}
\hline
$J_\varphi$ & $\amp$ & $t$ & $\theta_{z0}$ \\
\hline
1444.7 & 0.096 & 0.265 & $0.857$ \\
1502.6 & 0.055 & 0.365 & $-2.036$ \\
1560.5 & 0.105 & 0.414 & $1.014$ \\
1618.4 & 0.206 & 0.233 & $2.385$ \\
1676.3 & 0.274 & 0.217 & $-2.279$ \\
1734.2 & 0.195 & 0.260 & $1.859$ \\
1792.1 & 0.159 & 0.372 & $-2.530$ \\
1850.0 & 0.192 & 0.413 & $2.756$ \\
1907.9 & 0.129 & 0.467 & $0.803$ \\
1965.8 & 0.183 & 0.357 & $-0.590$ \\
2023.7 & 0.303 & 0.375 & $-0.721$ \\
2081.6 & 0.218 & 0.371 & $0.175$ \\
2139.5 & 0.243 & 0.416 & $-0.527$ \\
2197.4 & 0.284 & 0.463 & $-1.151$ \\
\hline
\end{tabular}
\end{table}

\subsection{Snail as a function of $J_\varphi, J_R$}\label{section:snail_JphiJR}

\begin{figure}
    \centering
    \includegraphics[width=\columnwidth]{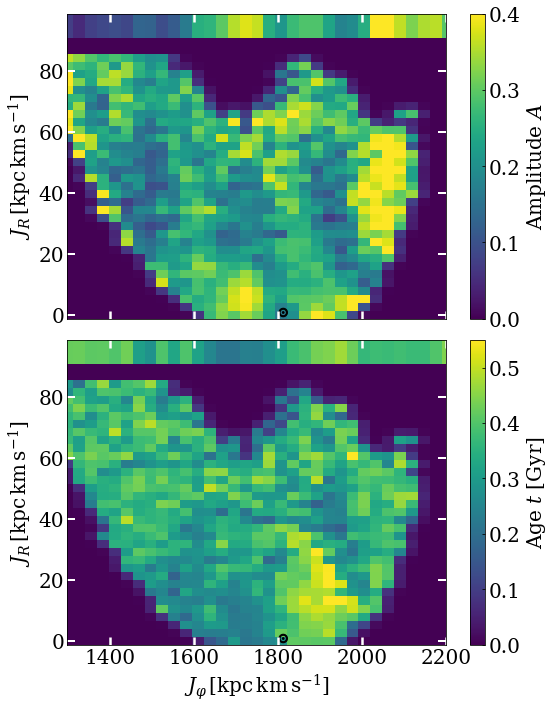}
    \caption{Parameters of the Snails inferred in bins of angular momentum and radial action. Only boxels with more than 800 stars are plotted. The top panel displays the best-fit Snail amplitude $\amp$ and the bottom panel shows the Snail age parameter $t$. The colored horizontal bands at the top of each panel are 1D representations of the results in Fig. \ref{fig:best_fit_trends}: In the top panel, the horizontal band displays $\amp (J_\varphi)\times 1.4$ to ease matching the variations by eye, while in the bottom panel the band shows $t(J_\varphi)$.}
    \label{fig:best_fit_JphiJz}
\end{figure}
Figs.\ \ref{fig:best_fit_trends} and \ref{fig:best_fit_trends_res} demonstrate that the Snail has significant trends with angular momentum $J_\varphi$ that are tightly constrained by the data. We now explore how the the parameters describing the Snail also vary with radial action $J_R$. Because we do not have any intuition on how the model parameters $\amp$, $t$ and $\theta_{z0}$ vary with $J_R$, we simply split the data in ${J_\varphi, J_R}$ bins and fit for the parameters in each bin, repeating the procedure described in \S\ref{section:model}. The best-fit values for the first two of these parameters are shown in Fig.\ \ref{fig:best_fit_JphiJz}. We only plot the bins that contain more than 800 stars (before the selection reweighting bootstrap).

The top panel of Fig.\ \ref{fig:best_fit_JphiJz} shows the best-fit amplitude as a function of $J_\varphi$ and $J_R$. First we note that in the top left region, $J_R \gtrsim 40\,\mathrm{kpc\,km\,s^{-1}}$, $J_\varphi \lesssim 1\,800\,\mathrm{kpc\,km\,s^{-1}}$, the signal is mostly dominated by statistical noise and it is hard to extract any specific trends. A visual check confirms that there is no visible Snail in $z$--$v_z$ space in this region of action space.
For $J_R \lesssim 40\,\mathrm{kpc\,km\,s^{-1}}$, we see with increasing $J_\varphi$ three yellow (high-amplitude) regions, near $J_\varphi \simeq 1\,730, 1\,900$ and $2\,050\, \mathrm{kpc\,km\,s}^{-1}$. The first two are strongest at low radial action, and disappear as $J_R$ increases above $10\,\mathrm{kpc\,km\,s}^{-1}$, while the third is centred near $J_R\simeq 40\,\mathrm{kpc\,km\,s}^{-1}$. In the top panel of Fig.\ \ref{fig:best_fit_trends} these regions correspond to three prominent peaks in the amplitude. The marginalized (over $J_R$) variations outlined in Fig.\ \ref{fig:best_fit_trends} are displayed as colored horizontal bands at the top of  Fig.\ \ref{fig:best_fit_JphiJz} to assist the visual comparisons. We also note that $\amp$ can occasionally take on large values at the edges of the coloured part of the $J_\varphi$--$J_R$ plane. 
We attribute the large amplitude values at low $J_\varphi$ to noisy estimates of $\amp$. Those at large $J_\varphi$ are physical and represent a strong Snail signal. 

The bottom panel shows the best-fit time parameter $t$. At low $J_\varphi$ where $\amp$ is small, the estimates of $t$ are noisy. In the region $1\,700 \lesssim J_\varphi \lesssim 1\,800\,\mathrm{kpc\,km\,s^{-1}}$, $t$ has its lowest values, $\simeq 0.2$ Gyr, at low radial actions.This low value matches the dip in Fig.\ \ref{fig:best_fit_trends}. Finally, the yellow area at $1\,800 \lesssim J_\varphi \lesssim 1\,980\,\mathrm{kpc\,km\,s^{-1}}$, has $t \simeq 0.5$--$0.6$ Gyr. The rapid transition from low to high $t$ matches that of Fig.\ \ref{fig:best_fit_trends}. In summary, the best-fit parameters $A$ and $t$ depend mostly on $J_\varphi$ and have only a modest dependence on $J_R$.

%
%
%
%
%
%
\section{Discussion \& Summary}\label{section:discussion}

We have shown that the properties of the \textsl{Gaia} Snail--and by extension, other phase-wrapped disturbances excited in the distant past--can be explored by plotting the row-normalized density of stars in the vertical frequency vs.\ vertical angle plane (the zebra diagram), after correcting for selection effects. In this plane, a single Fourier component of an instantaneous perturbation that is only weakly dependent on the action appears as a nearly straight line, with a slope that encapsulates the age of the perturbation. 

We have constructed a simple model for the Snail with three parameters: an amplitude $\amp$, a degree of winding represented by a parameter $t$ with the dimension of time, and a phase. We have fit this model to \textsl{Gaia} EDR3 data after accounting for selection effects as described in \S\ref{section:selection_function_correction}. 
The model parameters and trends were extracted for \textsl{Gaia} data in the solar neighbourhood (median distance from the Sun of 0.33 kpc), but the model could also be used to interpret the results from N-body simulations and compare them to the observational data.
We have found that the model parameters depend strongly on angular momentum $J_\varphi$, and modestly on radial action $J_R$. In particular, the amplitude shows a wave-like pattern with angular momentum (top panel of Fig.\ \ref{fig:best_fit_trends} which seems qualitatively similar to those found in $\langle v_z \rangle(J_\varphi)$ \citep{schoenrich_dehnen_2018} and in $\langle v_R\rangle (J_\varphi) $ \citep{friske_schoenrich_2019}, the latter being qualitatively compatible with  external perturbations that wind up \citep{antoja_2022}. This suggests that at least some of the vertical and in-plane deviations from a steady-state axisymmetric galaxy could be related.

If the Snail is excited by a single impulse, the parameter $t$ would be interpreted as the time elapsed since the impulse (the perturbation age), and $\amp$ would be a function of the mass of the perturber and the distance and relative velocity of the perturber at the point of closest approach. We found  $t\simeq 200$--$600$ Myr, which is consistent with the range of current literature values \citep{antoja_2018, laporte_2019_sgrfootprint, li_widrow_2021}. However, the variation in the time as a function of angular momentum
(top panel of Fig.\ \ref{fig:best_fit_trends}) is much larger than the range expected from a single impact, even allowing for variations in the time of closest approach with position in the disc
\citep{gandhi_2022_snail}. This behavior is also noted in \cite{Antoja_2022a}. We conclude that single-impulse models of the Snail are oversimplified, and the parameters derived here should be regarded as summary statistics whose relation to the physics of the excitation of the Snail is likely to depend on the excitation mechanism. 

A second problem with single-impulse theories is that the structure in the residuals is correlated, which indicates that the model cannot fully describe the data. The most prominent of these correlated residuals are $m=2$ modes at $J_\varphi \lesssim 1\,500 \mathrm{\, kpc\,km\,s^{-1}}$, reminiscent of the two-armed spiral described by \cite{hunt_2022}.

Our model ignores the effects of self-gravity in the evolution of the Snail. \cite{darling_widrow_2019} have argued that the amplitude and winding rate of the Snail are strongly affected by self-gravity, and this possibility deserves further exploration. However, once the Snail is tightly wound the corresponding density fluctuations--essentially the integration of the phase-space density over velocity $v_z$--will be small enough that they are unlikely to produce significant perturbations to the self-gravity. 

We have assumed that the Snail evolves in a fixed gravitational potential, and have assumed a specific form for this potential (\textsc{MWPotential2014} from \textsc{galpy}). We have tested the robustness of the results to the second of these assumptions using the \textsc{galpy} implementation of the Milky Way potential of \cite{mcmillan_2017}, and found similar results with typical differences in $\amp$ and $t$ of $\sim 0.03\,\mathrm{and}\,50\,\mathrm{Myr}$ respectively These are similar to the variations between different realizations of a model with fixed parameters, as described in in Section \ref{section:snail_Jphi}. The largest difference is that with the \cite{mcmillan_2017} potential, the drop in $t$ at $J_\varphi\simeq 2\,000\,\mathrm{kpc\,km\,s^{-1}}$ is more pronounced. An alternative approach would be to fit the potential self-consistently using the observed
distribution function \citep{widmark_2021}. 

We can generalize our model to multiple events and multiple azimuthal wavenumbers. In this case the analog of Eq.\ (\ref{eq:model_ml}) becomes 
\begin{align}
    p_\mathrm{MW}&(J_z,\theta_z| J_\varphi, J_R)=\frac{1}{2\pi}p_0(J_{z})\nonumber\\ &\times\sum_{i=1}^{N}\sum_{m=1}^M \amp_{im}\cos[m(\theta_z-\Omega_zt_{i}-\theta_{z0,im})],
\end{align}
The limit $N\gg1$ is described by \cite{tre22}. 

We have briefly explored the case of a single event ($N=1$) with two modes ($M=2$). We found that the best-fit $m=1$ parameters agree with the results presented in this paper. The $m=2$ mode has an amplitude that is strongest at low angular momentum, consistent with the residuals shown in Fig.\ \ref{fig:best_fit_trends_res}. However, it will be important in future work to account more carefully for the effects of dust extinction, which also produces $m=2$ effects due to the symmetry around the mid-plane of the Galactic disc. We also explored the case of two events ($N=2$) and a single mode ($M=1$). We found that in this case our fitting methods did not work well, as there were multiple local maxima of the likelihood and strong covariances between the parameters. Adding increasingly more events is interesting but prone to over-fitting any feature (physical or artifact) in the $\Omega_z$--$\theta_z$ plane, especially if other effects are such as dust and observational selection are not properly accounted for.

The model could be extended with other dimensions now available in the data; e.g., stellar age. Stars of greater and greater age may have witnessed more and more perturbation events. We note that \cite{bland-hawthorn_2019} have split a data set from the \textsc{galah} survey \citep{buder_2018} in chemistry space, which to some degree reflects age; adding this dimension could provide information about the physical timing of the perturbation events that is independent of inferences about timing from the degree of winding of the Snail.

We have made conservative cuts to the \textsl{Gaia} data, which require that we analyze only a small volume of the disc.  Increasing the survey volume may help to better quantify how the Snail parameters vary as a function of orbital actions and angles. Analyzing a larger volume will require (i) treating uncertainties in the measured stellar properties carefully and incorporating them in the likelihood function; (ii)  accounting for the effects of extinction from dust; (iii) accounting for effects from different stellar populations--since young stars are brighter and metal-rich stars redder, \textsl{Gaia} probes different stellar ages and birth sites as a function of position in the Galaxy. An interesting possibility is to restrict ourselves to a specific sub-population in a narrow region of the colour-magnitude diagram, such as red giant stars.

In conclusion, we have shown that the Snail can be quantified with a simple parameterized model that depends on the dynamical invariants or actions ($J_\varphi, J_z$). The resulting best-fit model implies that there is no well-defined global dynamical age of a single perturbation, and that the perturbation amplitude is a strong function of $J_\varphi$. Several extensions of this work, such as modelling additional dynamical invariants (age, [Fe/H]) or enlarging the data volume, have been made possible with the latest data release of Gaia but will require technical improvements in our methods, to be reserved for future work.

\section*{Acknowledgements}
It is a pleasure to thank Rimpei Chiba for stimulating discussions and a read of this paper. We thank Jason Hunt for interesting discussions.
This work has made use of data from the European Space Agency (ESA) mission \textsl{Gaia} (\url{https://www.cosmos.esa.int/gaia}), processed by the \textsl{Gaia} Data Processing and Analysis Consortium (DPAC, \url{https://www.cosmos.esa.int/web/gaia/dpac/consortium}). Funding for the DPAC has been provided by national institutions, in particular the institutions participating in the \textsl{Gaia} Multilateral Agreement.
NF was supported by the Natural Sciences and Engineering Research Council of Canada (NSERC), funding reference number CITA 490888-16, through a CITA postdoctoral fellowship, and acknowledges partial support from an Arts \& Sciences Postdoctoral Fellowship at the University of Toronto. JB and ST also received support from NSERC, funding references RGPIN-2020-04712 and RGPIN-2020-03885.
This research was supported in part by the National Science Foundation under Grant No. NSF PHY-1748958. NF is grateful to Hans-Walter Rix and the MPIA for providing office space, hosting part of this research in the Hammock Under The Tree B\"uro.

\section*{Data availability}
The observational data underlying this article were accessed from
the Gaia archive (https://gea.esac.esa.int/archive/). The data underlying this article will be shared on reasonable request to the corresponding author.

\bibliographystyle{mnras}
\bibliography{lit}

\begin{thebibliography}{}
\makeatletter
\relax
\def\mn@urlcharsother{\let\do\@makeother \do\$\do\&\do\#\do\^\do\_\do\%\do\~}
\def\mn@doi{\begingroup\mn@urlcharsother \@ifnextchar [ {\mn@doi@}
  {\mn@doi@[]}}
\def\mn@doi@[#1]#2{\def\@tempa{#1}\ifx\@tempa\@empty \href
  {http://dx.doi.org/#2} {doi:#2}\else \href {http://dx.doi.org/#2} {#1}\fi
  \endgroup}
\def\mn@eprint#1#2{\mn@eprint@#1:#2::\@nil}
\def\mn@eprint@arXiv#1{\href {http://arxiv.org/abs/#1} {{\tt arXiv:#1}}}
\def\mn@eprint@dblp#1{\href {http://dblp.uni-trier.de/rec/bibtex/#1.xml}
  {dblp:#1}}
\def\mn@eprint@#1:#2:#3:#4\@nil{\def\@tempa {#1}\def\@tempb {#2}\def\@tempc
  {#3}\ifx \@tempc \@empty \let \@tempc \@tempb \let \@tempb \@tempa \fi \ifx
  \@tempb \@empty \def\@tempb {arXiv}\fi \@ifundefined
  {mn@eprint@\@tempb}{\@tempb:\@tempc}{\expandafter \expandafter \csname
  mn@eprint@\@tempb\endcsname \expandafter{\@tempc}}}

\bibitem[\protect\citeauthoryear{{Antoja} et~al.,}{{Antoja}
  et~al.}{2018}]{antoja_2018}
{Antoja} T.,  et~al., 2018, \mn@doi [\nat] {10.1038/s41586-018-0510-7}, \href
  {https://ui.adsabs.harvard.edu/abs/2018Natur.561..360A} {561, 360}

\bibitem[\protect\citeauthoryear{{Antoja}, {Ramos}, {Garcia-Conde}, {Bernet},
  {Laporte}  \& {Katz}}{{Antoja} et~al.}{2022a}]{Antoja_2022a}
{Antoja} T.,  {Ramos} P.,  {Garcia-Conde} B.,  {Bernet} M.,  {Laporte} C.,
  {Katz} D.,  2022a, submitted to \aap

\bibitem[\protect\citeauthoryear{{Antoja}, {Ramos}, {L{\'o}pez-Guitart},
  {Anders}, {Bernet}  \& {Laporte}}{{Antoja} et~al.}{2022b}]{antoja_2022}
{Antoja} T.,  {Ramos} P.,  {L{\'o}pez-Guitart} F.,  {Anders} F.,  {Bernet} M.,
   {Laporte} C.~F.~P.,  2022b, \mn@doi [\aap] {10.1051/0004-6361/202244064},
  \href {https://ui.adsabs.harvard.edu/abs/2022A&A...668A..61A} {668, A61}

\bibitem[\protect\citeauthoryear{{Banik}, {Weinberg}  \& {van den
  Bosch}}{{Banik} et~al.}{2022}]{banik_2022}
{Banik} U.,  {Weinberg} M.~D.,   {van den Bosch} F.~C.,  2022, \mn@doi [\apj]
  {10.3847/1538-4357/ac7ff9}, \href
  {https://ui.adsabs.harvard.edu/abs/2022ApJ...935..135B} {935, 135}

\bibitem[\protect\citeauthoryear{{Bennett} \& {Bovy}}{{Bennett} \&
  {Bovy}}{2019}]{bennett_bovy_2019}
{Bennett} M.,  {Bovy} J.,  2019, \mn@doi [\mnras] {10.1093/mnras/sty2813},
  \href {https://ui.adsabs.harvard.edu/abs/2019MNRAS.482.1417B} {482, 1417}

\bibitem[\protect\citeauthoryear{{Bennett} \& {Bovy}}{{Bennett} \&
  {Bovy}}{2021}]{bennett_bovy_2021}
{Bennett} M.,  {Bovy} J.,  2021, \mn@doi [\mnras] {10.1093/mnras/stab524},
  \href {https://ui.adsabs.harvard.edu/abs/2021MNRAS.503..376B} {503, 376}

\bibitem[\protect\citeauthoryear{{Bennett}, {Bovy}  \& {Hunt}}{{Bennett}
  et~al.}{2022}]{bennett_2022}
{Bennett} M.,  {Bovy} J.,   {Hunt} J. A.~S.,  2022, \mn@doi [\apj]
  {10.3847/1538-4357/ac5021}, \href
  {https://ui.adsabs.harvard.edu/abs/2022ApJ...927..131B} {927, 131}

\bibitem[\protect\citeauthoryear{{Binney}}{{Binney}}{2012}]{binney_2012}
{Binney} J.,  2012, \mn@doi [\mnras] {10.1111/j.1365-2966.2012.21757.x}, \href
  {https://ui.adsabs.harvard.edu/abs/2012MNRAS.426.1324B} {426, 1324}

\bibitem[\protect\citeauthoryear{{Bland-Hawthorn} et~al.,}{{Bland-Hawthorn}
  et~al.}{2019}]{bland-hawthorn_2019}
{Bland-Hawthorn} J.,  et~al., 2019, \mn@doi [\mnras] {10.1093/mnras/stz217},
  \href {https://ui.adsabs.harvard.edu/abs/2019MNRAS.486.1167B} {486, 1167}

\bibitem[\protect\citeauthoryear{{Bovy}}{{Bovy}}{2015}]{bovy_2015_galpy}
{Bovy} J.,  2015, \mn@doi [\apjs] {10.1088/0067-0049/216/2/29}, \href
  {https://ui.adsabs.harvard.edu/abs/2015ApJS..216...29B} {216, 29}

\bibitem[\protect\citeauthoryear{{Bovy}, {Rix}, {Liu}, {Hogg}, {Beers}  \&
  {Lee}}{{Bovy} et~al.}{2012}]{bovy_etal_2012}
{Bovy} J.,  {Rix} H.-W.,  {Liu} C.,  {Hogg} D.~W.,  {Beers} T.~C.,   {Lee}
  Y.~S.,  2012, \mn@doi [\apj] {10.1088/0004-637X/753/2/148}, \href
  {http://cdsads.u-strasbg.fr/abs/2012ApJ...753..148B} {753, 148}

\bibitem[\protect\citeauthoryear{{Bovy}, {Rix}, {Schlafly}, {Nidever},
  {Holtzman}, {Shetrone}  \& {Beers}}{{Bovy} et~al.}{2016}]{bovy_etal_2016}
{Bovy} J.,  {Rix} H.-W.,  {Schlafly} E.~F.,  {Nidever} D.~L.,  {Holtzman}
  J.~A.,  {Shetrone} M.,   {Beers} T.~C.,  2016, \mn@doi [\apj]
  {10.3847/0004-637X/823/1/30}, \href
  {http://adsabs.harvard.edu/abs/2016ApJ...823...30B} {823, 30}

\bibitem[\protect\citeauthoryear{{Buder} et~al.,}{{Buder}
  et~al.}{2018}]{buder_2018}
{Buder} S.,  et~al., 2018, \mn@doi [\mnras] {10.1093/mnras/sty1281}, \href
  {https://ui.adsabs.harvard.edu/abs/2018MNRAS.478.4513B} {478, 4513}

\bibitem[\protect\citeauthoryear{{Darling} \& {Widrow}}{{Darling} \&
  {Widrow}}{2019}]{darling_widrow_2019}
{Darling} K.,  {Widrow} L.~M.,  2019, \mn@doi [\mnras] {10.1093/mnras/sty3508},
  \href {https://ui.adsabs.harvard.edu/abs/2019MNRAS.484.1050D} {484, 1050}

\bibitem[\protect\citeauthoryear{{Erkal} et~al.,}{{Erkal}
  et~al.}{2019}]{erkal_2019}
{Erkal} D.,  et~al., 2019, \mn@doi [\mnras] {10.1093/mnras/stz1371}, \href
  {https://ui.adsabs.harvard.edu/abs/2019MNRAS.487.2685E} {487, 2685}

\bibitem[\protect\citeauthoryear{{Friske} \& {Sch{\"o}nrich}}{{Friske} \&
  {Sch{\"o}nrich}}{2019}]{friske_schoenrich_2019}
{Friske} J. K.~S.,  {Sch{\"o}nrich} R.,  2019, \mn@doi [\mnras]
  {10.1093/mnras/stz2951}, \href
  {https://ui.adsabs.harvard.edu/abs/2019MNRAS.490.5414F} {490, 5414}

\bibitem[\protect\citeauthoryear{{Gaia Collaboration} et~al.,}{{Gaia
  Collaboration} et~al.}{2016}]{gaia_prusti_2016}
{Gaia Collaboration} et~al., 2016, \mn@doi [\aap]
  {10.1051/0004-6361/201629272}, \href
  {https://ui.adsabs.harvard.edu/abs/2016A&A...595A...1G} {595, A1}

\bibitem[\protect\citeauthoryear{{Gaia Collaboration} et~al.,}{{Gaia
  Collaboration} et~al.}{2018}]{gaia_dr2_2018}
{Gaia Collaboration} et~al., 2018, \mn@doi [\aap]
  {10.1051/0004-6361/201833051}, \href
  {https://ui.adsabs.harvard.edu/abs/2018A&A...616A...1G} {616, A1}

\bibitem[\protect\citeauthoryear{{Gaia Collaboration} et~al.,}{{Gaia
  Collaboration} et~al.}{2021}]{Gaia_EDR3_Brown_2021}
{Gaia Collaboration} et~al., 2021, \mn@doi [\aap]
  {10.1051/0004-6361/202039657}, \href
  {https://ui.adsabs.harvard.edu/abs/2021A&A...649A...1G} {649, A1}

\bibitem[\protect\citeauthoryear{{Gandhi}, {Johnston}, {Hunt}, {Price-Whelan},
  {Laporte}  \& {Hogg}}{{Gandhi} et~al.}{2022}]{gandhi_2022_snail}
{Gandhi} S.~S.,  {Johnston} K.~V.,  {Hunt} J. A.~S.,  {Price-Whelan} A.~M.,
  {Laporte} C. F.~P.,   {Hogg} D.~W.,  2022, \mn@doi [\apj]
  {10.3847/1538-4357/ac47f7}, \href
  {https://ui.adsabs.harvard.edu/abs/2022ApJ...928...80G} {928, 80}

\bibitem[\protect\citeauthoryear{{Garc{\'\i}a-Conde}, {Roca-F{\`a}brega},
  {Antoja}, {Ramos}  \& {Valenzuela}}{{Garc{\'\i}a-Conde}
  et~al.}{2022}]{garcia_conde_2022}
{Garc{\'\i}a-Conde} B.,  {Roca-F{\`a}brega} S.,  {Antoja} T.,  {Ramos} P.,
  {Valenzuela} O.,  2022, \mn@doi [\mnras] {10.1093/mnras/stab3417}, \href
  {https://ui.adsabs.harvard.edu/abs/2022MNRAS.510..154G} {510, 154}

\bibitem[\protect\citeauthoryear{{Grand}, {Pakmor}, {Fragkoudi}, {G{\'o}mez},
  {Trick}, {Simpson}, {van de Voort}  \& {Bieri}}{{Grand}
  et~al.}{2022}]{grand_2022}
{Grand} R. J.~J.,  {Pakmor} R.,  {Fragkoudi} F.,  {G{\'o}mez} F.~A.,  {Trick}
  W.,  {Simpson} C.~M.,  {van de Voort} F.,   {Bieri} R.,  2022, arXiv
  e-prints, \href {https://ui.adsabs.harvard.edu/abs/2022arXiv221108437G} {p.
  arXiv:2211.08437}

\bibitem[\protect\citeauthoryear{{Hunt}, {Stelea}, {Johnston}, {Gandhi},
  {Laporte}  \& {B{\'e}dorf}}{{Hunt} et~al.}{2021}]{hunt_2021}
{Hunt} J. A.~S.,  {Stelea} I.~A.,  {Johnston} K.~V.,  {Gandhi} S.~S.,
  {Laporte} C. F.~P.,   {B{\'e}dorf} J.,  2021, \mn@doi [\mnras]
  {10.1093/mnras/stab2580}, \href
  {https://ui.adsabs.harvard.edu/abs/2021MNRAS.508.1459H} {508, 1459}

\bibitem[\protect\citeauthoryear{{Hunt}, {Price-Whelan}, {Johnston}  \&
  {Darragh-Ford}}{{Hunt} et~al.}{2022}]{hunt_2022}
{Hunt} J. A.~S.,  {Price-Whelan} A.~M.,  {Johnston} K.~V.,   {Darragh-Ford} E.,
   2022, \mn@doi [\mnras] {10.1093/mnrasl/slac082}, \href
  {https://ui.adsabs.harvard.edu/abs/2022MNRAS.516L...7H} {516, L7}

\bibitem[\protect\citeauthoryear{{Ibata}, {Gilmore}  \& {Irwin}}{{Ibata}
  et~al.}{1994}]{ibata_1994}
{Ibata} R.~A.,  {Gilmore} G.,   {Irwin} M.~J.,  1994, \mn@doi [\nat]
  {10.1038/370194a0}, \href
  {https://ui.adsabs.harvard.edu/abs/1994Natur.370..194I} {370, 194}

\bibitem[\protect\citeauthoryear{{Ibata}, {Wyse}, {Gilmore}, {Irwin}  \&
  {Suntzeff}}{{Ibata} et~al.}{1997}]{ibata_1997}
{Ibata} R.~A.,  {Wyse} R. F.~G.,  {Gilmore} G.,  {Irwin} M.~J.,   {Suntzeff}
  N.~B.,  1997, \mn@doi [\aj] {10.1086/118283}, \href
  {https://ui.adsabs.harvard.edu/abs/1997AJ....113..634I} {113, 634}

\bibitem[\protect\citeauthoryear{{Khoperskov}, {Di Matteo}, {Gerhard}, {Katz},
  {Haywood}, {Combes}, {Berczik}  \& {Gomez}}{{Khoperskov}
  et~al.}{2019}]{khoperskov_2019}
{Khoperskov} S.,  {Di Matteo} P.,  {Gerhard} O.,  {Katz} D.,  {Haywood} M.,
  {Combes} F.,  {Berczik} P.,   {Gomez} A.,  2019, \mn@doi [\aap]
  {10.1051/0004-6361/201834707}, \href
  {https://ui.adsabs.harvard.edu/abs/2019A&A...622L...6K} {622, L6}

\bibitem[\protect\citeauthoryear{{Kormendy} \& {Kennicutt}}{{Kormendy} \&
  {Kennicutt}}{2004}]{kormendy_2004A}
{Kormendy} J.,  {Kennicutt} Robert~C. J.,  2004, \mn@doi [\araa]
  {10.1146/annurev.astro.42.053102.134024}, \href
  {https://ui.adsabs.harvard.edu/abs/2004ARA&A..42..603K} {42, 603}

\bibitem[\protect\citeauthoryear{{Laporte}, {Johnston}, {G{\'o}mez},
  {Garavito-Camargo}  \& {Besla}}{{Laporte} et~al.}{2018}]{laporte_2018_sgr}
{Laporte} C. F.~P.,  {Johnston} K.~V.,  {G{\'o}mez} F.~A.,  {Garavito-Camargo}
  N.,   {Besla} G.,  2018, \mn@doi [\mnras] {10.1093/mnras/sty1574}, \href
  {https://ui.adsabs.harvard.edu/abs/2018MNRAS.481..286L} {481, 286}

\bibitem[\protect\citeauthoryear{{Laporte}, {Minchev}, {Johnston}  \&
  {G{\'o}mez}}{{Laporte} et~al.}{2019}]{laporte_2019_sgrfootprint}
{Laporte} C. F.~P.,  {Minchev} I.,  {Johnston} K.~V.,   {G{\'o}mez} F.~A.,
  2019, \mn@doi [\mnras] {10.1093/mnras/stz583}, \href
  {https://ui.adsabs.harvard.edu/abs/2019MNRAS.485.3134L} {485, 3134}

\bibitem[\protect\citeauthoryear{{Leung}, {Bovy}, {Mackereth}, {Hunt}, {Lane}
  \& {Wilson}}{{Leung} et~al.}{2022}]{leung_2022}
{Leung} H.~W.,  {Bovy} J.,  {Mackereth} J.~T.,  {Hunt} J. A.~S.,  {Lane} R.~R.,
    {Wilson} J.~C.,  2022, arXiv e-prints, \href
  {https://ui.adsabs.harvard.edu/abs/2022arXiv220412551L} {p. arXiv:2204.12551}

\bibitem[\protect\citeauthoryear{{Li} \& {Shen}}{{Li} \& {Shen}}{2020}]{ls20}
{Li} Z.-Y.,  {Shen} J.,  2020, \mn@doi [\apj] {10.3847/1538-4357/ab6b21}, \href
  {https://ui.adsabs.harvard.edu/abs/2020ApJ...890...85L} {890, 85}

\bibitem[\protect\citeauthoryear{{Li} \& {Widrow}}{{Li} \&
  {Widrow}}{2021}]{li_widrow_2021}
{Li} H.,  {Widrow} L.~M.,  2021, \mn@doi [\mnras] {10.1093/mnras/stab574},
  \href {https://ui.adsabs.harvard.edu/abs/2021MNRAS.503.1586L} {503, 1586}

\bibitem[\protect\citeauthoryear{{Lynden-Bell} \& {Kalnajs}}{{Lynden-Bell} \&
  {Kalnajs}}{1972}]{lynden-bell_kalnajs_1972}
{Lynden-Bell} D.,  {Kalnajs} A.~J.,  1972, \mn@doi [\mnras]
  {10.1093/mnras/157.1.1}, \href
  {https://ui.adsabs.harvard.edu/abs/1972MNRAS.157....1L} {157, 1}

\bibitem[\protect\citeauthoryear{{McMillan}}{{McMillan}}{2017}]{mcmillan_2017}
{McMillan} P.~J.,  2017, \mn@doi [\mnras] {10.1093/mnras/stw2759}, \href
  {https://ui.adsabs.harvard.edu/abs/2017MNRAS.465...76M} {465, 76}

\bibitem[\protect\citeauthoryear{{Mo}, {Mao}  \& {White}}{{Mo}
  et~al.}{1998}]{mo_mao_white_1998}
{Mo} H.~J.,  {Mao} S.,   {White} S.~D.~M.,  1998, \mn@doi [\mnras]
  {10.1046/j.1365-8711.1998.01227.x}, \href
  {http://cdsads.u-strasbg.fr/abs/1998MNRAS.295..319M} {295, 319}

\bibitem[\protect\citeauthoryear{Nelder \& Mead}{Nelder \&
  Mead}{1965}]{NeldMead65}
Nelder J.~A.,  Mead R.,  1965, Computer Journal, 7, 308

\bibitem[\protect\citeauthoryear{{Pietrzy{\'n}ski} et~al.,}{{Pietrzy{\'n}ski}
  et~al.}{2013}]{pietrzynski_2013}
{Pietrzy{\'n}ski} G.,  et~al., 2013, \mn@doi [\nat] {10.1038/nature11878},
  \href {https://ui.adsabs.harvard.edu/abs/2013Natur.495...76P} {495, 76}

\bibitem[\protect\citeauthoryear{{Poggio}, {Laporte}, {Johnston}, {D'Onghia},
  {Drimmel}  \& {Grion Filho}}{{Poggio} et~al.}{2021}]{poggio_2021}
{Poggio} E.,  {Laporte} C. F.~P.,  {Johnston} K.~V.,  {D'Onghia} E.,  {Drimmel}
  R.,   {Grion Filho} D.,  2021, \mn@doi [\mnras] {10.1093/mnras/stab2245},
  \href {https://ui.adsabs.harvard.edu/abs/2021MNRAS.508..541P} {508, 541}

\bibitem[\protect\citeauthoryear{{Rees} \& {Ostriker}}{{Rees} \&
  {Ostriker}}{1977}]{rees_ostriker_1977}
{Rees} M.~J.,  {Ostriker} J.~P.,  1977, \mn@doi [\mnras]
  {10.1093/mnras/179.4.541}, \href
  {https://ui.adsabs.harvard.edu/abs/1977MNRAS.179..541R} {179, 541}

\bibitem[\protect\citeauthoryear{{Rybizki}, {Rix}, {Demleitner}, {Bailer-Jones}
   \& {Cooper}}{{Rybizki} et~al.}{2021}]{rybizki_2021}
{Rybizki} J.,  {Rix} H.-W.,  {Demleitner} M.,  {Bailer-Jones} C. A.~L.,
  {Cooper} W.~J.,  2021, \mn@doi [\mnras] {10.1093/mnras/staa3089}, \href
  {https://ui.adsabs.harvard.edu/abs/2021MNRAS.500..397R} {500, 397}

\bibitem[\protect\citeauthoryear{{Sch{\"o}nrich} \& {Dehnen}}{{Sch{\"o}nrich}
  \& {Dehnen}}{2018}]{schoenrich_dehnen_2018}
{Sch{\"o}nrich} R.,  {Dehnen} W.,  2018, \mn@doi [\mnras]
  {10.1093/mnras/sty1256}, \href
  {https://ui.adsabs.harvard.edu/abs/2018MNRAS.478.3809S} {478, 3809}

\bibitem[\protect\citeauthoryear{{Sch{\"o}nrich}, {Binney}  \&
  {Dehnen}}{{Sch{\"o}nrich} et~al.}{2010}]{schoenrich_2010}
{Sch{\"o}nrich} R.,  {Binney} J.,   {Dehnen} W.,  2010, \mn@doi [\mnras]
  {10.1111/j.1365-2966.2010.16253.x}, \href
  {https://ui.adsabs.harvard.edu/abs/2010MNRAS.403.1829S} {403, 1829}

\bibitem[\protect\citeauthoryear{{Scott}}{{Scott}}{1992}]{scott_KDEBandwidth}
{Scott} D.~W.,  1992, {Multivariate Density Estimation}

\bibitem[\protect\citeauthoryear{{Sellwood} \& {Binney}}{{Sellwood} \&
  {Binney}}{2002}]{sellwood_binney_2002}
{Sellwood} J.~A.,  {Binney} J.~J.,  2002, \mn@doi [\mnras]
  {10.1046/j.1365-8711.2002.05806.x}, \href
  {http://cdsads.u-strasbg.fr/abs/2002MNRAS.336..785S} {336, 785}

\bibitem[\protect\citeauthoryear{{Tepper-Garcia} et~al.,}{{Tepper-Garcia}
  et~al.}{2021}]{tepper21}
{Tepper-Garcia} T.,  et~al., 2021, arXiv e-prints, \href
  {https://ui.adsabs.harvard.edu/abs/2021arXiv211105466T} {p. arXiv:2111.05466}

\bibitem[\protect\citeauthoryear{{Ting}, {Conroy}, {Rix}  \& {Cargile}}{{Ting}
  et~al.}{2019}]{ting_2019}
{Ting} Y.-S.,  {Conroy} C.,  {Rix} H.-W.,   {Cargile} P.,  2019, \mn@doi [\apj]
  {10.3847/1538-4357/ab2331}, \href
  {https://ui.adsabs.harvard.edu/abs/2019ApJ...879...69T} {879, 69}

\bibitem[\protect\citeauthoryear{{Tremaine}, {Frankel}  \& {Bovy}}{{Tremaine}
  et~al.}{2022}]{tre22}
{Tremaine} S.,  {Frankel} N.,   {Bovy} J.,  2022, submitted to \mnras

\bibitem[\protect\citeauthoryear{{Vasiliev} \& {Belokurov}}{{Vasiliev} \&
  {Belokurov}}{2020}]{vasiliev_belokurov_2020}
{Vasiliev} E.,  {Belokurov} V.,  2020, \mn@doi [\mnras]
  {10.1093/mnras/staa2114}, \href
  {https://ui.adsabs.harvard.edu/abs/2020MNRAS.497.4162V} {497, 4162}

\bibitem[\protect\citeauthoryear{{Vasiliev}, {Belokurov}  \&
  {Erkal}}{{Vasiliev} et~al.}{2021}]{vasiliev_2021}
{Vasiliev} E.,  {Belokurov} V.,   {Erkal} D.,  2021, \mn@doi [\mnras]
  {10.1093/mnras/staa3673}, \href
  {https://ui.adsabs.harvard.edu/abs/2021MNRAS.501.2279V} {501, 2279}

\bibitem[\protect\citeauthoryear{{Virtanen} et~al.,}{{Virtanen}
  et~al.}{2020}]{scipy}
{Virtanen} P.,  et~al., 2020, \mn@doi [Nature Methods]
  {https://doi.org/10.1038/s41592-019-0686-2}, \href {https://rdcu.be/b08Wh}
  {17, 261}

\bibitem[\protect\citeauthoryear{{White} \& {Rees}}{{White} \&
  {Rees}}{1978}]{white_rees_1978}
{White} S.~D.~M.,  {Rees} M.~J.,  1978, \mn@doi [\mnras]
  {10.1093/mnras/183.3.341}, \href
  {https://ui.adsabs.harvard.edu/abs/1978MNRAS.183..341W} {183, 341}

\bibitem[\protect\citeauthoryear{{Widmark}, {Laporte}  \& {de Salas}}{{Widmark}
  et~al.}{2021}]{widmark_2021}
{Widmark} A.,  {Laporte} C.,   {de Salas} P.~F.,  2021, \mn@doi [\aap]
  {10.1051/0004-6361/202140650}, \href
  {https://ui.adsabs.harvard.edu/abs/2021A&A...650A.124W} {650, A124}

\bibitem[\protect\citeauthoryear{{Widrow}, {Gardner}, {Yanny}, {Dodelson}  \&
  {Chen}}{{Widrow} et~al.}{2012}]{widrow_2012}
{Widrow} L.~M.,  {Gardner} S.,  {Yanny} B.,  {Dodelson} S.,   {Chen} H.-Y.,
  2012, \mn@doi [\apjl] {10.1088/2041-8205/750/2/L41}, \href
  {https://ui.adsabs.harvard.edu/abs/2012ApJ...750L..41W} {750, L41}

\bibitem[\protect\citeauthoryear{{Widrow}, {Barber}, {Chequers}  \&
  {Cheng}}{{Widrow} et~al.}{2014}]{widrow_2014}
{Widrow} L.~M.,  {Barber} J.,  {Chequers} M.~H.,   {Cheng} E.,  2014, \mn@doi
  [\mnras] {10.1093/mnras/stu396}, \href
  {https://ui.adsabs.harvard.edu/abs/2014MNRAS.440.1971W} {440, 1971}

\bibitem[\protect\citeauthoryear{{Xu}, {Newberg}, {Carlin}, {Liu}, {Deng},
  {Li}, {Sch{\"o}nrich}  \& {Yanny}}{{Xu} et~al.}{2015}]{xu_2015}
{Xu} Y.,  {Newberg} H.~J.,  {Carlin} J.~L.,  {Liu} C.,  {Deng} L.,  {Li} J.,
  {Sch{\"o}nrich} R.,   {Yanny} B.,  2015, \mn@doi [\apj]
  {10.1088/0004-637X/801/2/105}, \href
  {https://ui.adsabs.harvard.edu/abs/2015ApJ...801..105X} {801, 105}

\makeatother
\end{thebibliography}

\appendix

\section{Data Query} \label{appendix:data_query}

The query used to produce the sample from \textsl{Gaia} EDR3 that is described in \S\ref{section:data_selection} and analyzed in this work is displayed below. 
\begin{verbatim}
SELECT
dr2_radial_velocity as radial_velocity, 
dr2_radial_velocity_error as radial_velocity_error, 
phot_g_mean_mag, bp_rp, g_rp,
ra, dec, parallax, parallax_error, pmra, 
pmra_error, pmdec, pmdec_error
FROM gaiaEDR3.gaia_source
WHERE dr2_radial_velocity IS NOT Null 
AND parallax IS NOT Null
AND g_rp < 1.25
AND g_rp > 0.35
AND phot_g_mean_mag < 12.5
\end{verbatim}

\label{lastpage}
\end{document}